\documentclass[11pt]{letter}
\usepackage{amsmath}
\usepackage{amssymb}
\usepackage[dvips]{graphicx}
\DeclareFontFamily{U}{msb}{}
\DeclareFontShape{U}{msb}{m}{n}{ <5> <6> <7> <8> <9> gen * msbm
        <10> <10.95> <12> <14.4> <17.28> <20.74> <24.88> msbm10}{}
\DeclareSymbolFont{AMSb}{U}{msb}{m}{n}
\DeclareMathSymbol{\realset}{\mathalpha}{AMSb}{"52}
\newcommand{\mbi}{\ensuremath{\mathbf{i}}}
\newcommand{\mbj}{\ensuremath{\mathbf{j}}}
\newcommand{\mbk}{\ensuremath{\mathbf{k}}}
\newcommand{\mbl}{\ensuremath{\mathbf{l}}}

\begin{document}

\begin{center}
\Large 
{\bf A Central Partition of Molecular Conformational Space. \newline
       III. Combinatorial Determination of the Volume Spanned 
            by a Molecular System.}
\end{center}

\vspace*{4mm}
\begin{center}
{\Large
Jacques Gabarro-Arpa}\footnote{Email: jga@infobiogen.fr}
\end{center}

\begin{center}
\hspace*{12mm}
Ecole Normale Sup\'erieure de Cachan, LBPA,CNRS UMR 8113  \newline
 61, Avenue du Pr\'esident Wilson, 94235 Cachan cedex, France
\end{center}

\vspace*{4mm}
\hspace*{4mm} {\it Abstract---} 
In the first work of this series [1] it was shown that the conformational 
space of a molecule could be described to a fair degree of accuracy 
by means of a central hyperplane arrangement.
The hyperplanes divide the espace into a hierarchical set of cells
that can be encoded by the face lattice poset of the arrangement. \newline
The model however, lacked explicit rotational symmetry which made impossible 
to distinguish rotated structures in conformational space. 
This problem was solved in a second work [2] by sorting the elementary 
$3D$ components of the molecular system into a set of morphological classes
that can be properly oriented in a standard $3D$ reference frame. \newline
This also made possible to find a solution to the problem that is being 
adressed in the present work:
{\it for a molecular system immersed in a heat bath we want to enumerate
 the subset of cells in conformational space that are visited by the molecule 
 in its thermal wandering}. \newline
If each visited cell is a vertex on a graph with edges to the adjacent cells,
here it is explained how such graph can be built.

\vspace*{2mm}
\hspace*{4mm} {\it Keywords---}
{\bf Molecular Conformational Space, Hyperplane Arrangement, Face Lattice,
     Molecular Dynamics } 



\vspace*{4mm}
\begin{center}
{\scshape I. Introduction}
\end{center}

Molecular dynamics simulations ($MDS$) are an essential tool for the 
modeling of large and very large molecules, it gives us a precise 
and detailed view of a molecule's behaviour [3]. However, it has 
two limitations that hamper many practical applications: it is a random 
algorithm, as such it does not perform a systematic exploration of molecular 
conformational space ($CS$); and that currently, the output from an $MDS$ 
represents only a very small fraction of the volume spanned by the system 
in $CS$.

\vskip 1mm
\noindent
Here it is presented a complementary approach that locally is less
precise but that can encompass a broader view of $CS$. It consists
in dividing the $CS$ into a finite set of cells, so that the only knowledge 
we seek about the system is whether it can be located in a given cell or not.

\vskip 1mm
\noindent
As was extensively discussed in ref. [1] the partition is a variant 
of the $\mathcal{A}_{N}$ partition [4-5]: 
a {\bf central}\footnote{That pass through the origin.} 
arrangement of hyperplanes that divides $CS$ into a set of cells shaped 
as polyhedral cones, such that for a molecule with $N$ atoms we have 
$(N!)^{\text{\small{3}}}$ cells. 
The set of hyperplanes is also a Coxeter reflection arrangement:
the arrangement is invariant upon reflection on any of the hyperplanes.
\vskip 1mm
\noindent
This structure has three important properties [1]:
\begin{enumerate}
\item Associated with a Coxeter arrangement there is a polytope [5] whose
      symmetry group is the reflection group of the arrangement.
      The {\bf face lattice poset}\footnote{The faces in the induced 
      decomposition of the polytope ordered by inclusion.}
      of the polytope is a hierarchichal combinatorial structure that 
      enables us to manage the sheer complexity of $CS$, since with simple
      codes we can describe from huge regions down to single cells.
\item The information needed to encode any face in the polytope is
      a sequence of $3 \times N$ integers, which is a generalization 
      of a structure known to combinatorialists as {\bf non-crossing partition 
      sequence} [5,6].
\item The construction is {\bf modular}: if we consider the $CS$ 
      of two subsets of atoms from a system, the $CS$ of the union set 
      has an associated polytope that is the cartesian product 
      of the polytopes\footnote{If $P \subset \realset^{p}$
      and $Q \subset \realset^{q}$ are polytopes the product polytope 
      $P \times Q$ has the set of vertices $(x,y) \ \epsilon \ \realset^{p+q}$
      where $x$ and $y$ are vertices of $P$ and $Q$ respectively.}
      of the two subspaces, and its partition sequence is the ordered union
      of the two partition sequences [1].
\end{enumerate}
The last one is particularly important since the $CS$ of the whole system 
can be built from that of the parts, and the $CS$ of a small number 
of atoms is very much smaller than that of the whole molecule 
and we can reasonably assume that it can be thoroughly explored by an $MDS$. 
Moreover, in merging the $CS$s corresponding to subsets of atoms the number 
of cells grows exponentially while the length of coding sequences 
grows only linearly.

\begin{center}
{\scshape II. The basic construction}
\end{center}

Let $(e_1, ... ,e_N)$ be the standard basis in $\realset^N$, 
the convex hull of the endpoints of the vectors $ \{ e_{\mbi} \} $ 
is a {\bf regular ${(N \! \! - \! \! 1)}$-simplex} : 
this gives a segment, an equilateral triangle and a tetrahedron in 2, 3 
and 4 dimensions respectively.

\par
For each edge of the regular ${(N \! \! - \! \! 1)}$-simplex there is 
an hyperplane $H_{\mbi\mbj}$ : $x_{\mbi}-x_{\mbj}=0$, 
perpendicular to the edge and containing the other vertices, this hyperplane 
divides $\realset^N$ in three regions. 
A point $x$ can be in one of these : 

\begin{itemize}
\item $x_{\mbi} > x_{\mbj}$ the positive side, 
      where the $\mbi^{\text{th}}$ coordinate {\bf dominates} 
     the $\mbj^{\text{th}}$ coordinate,
\item $x_{\mbi} < x_{\mbj}$ the negative side, 
      whith the $\mbj^{\text{th}}$ the coordinate dominating 
      the $\mbi^{\text{th}}$ coordinate,
\item $x_{\mbi} = x_{\mbj}$ on the plane.
\end{itemize}

\par
This leads to a sign vector $S$ for every point 
$x \ \epsilon \ \realset^N$, where the ${\alpha}^{\text{th}}$ component 
$X_{\alpha} \ \epsilon \ \{+,-,0\}$ denotes wether $x$ is
on the positive side of $H_{\alpha}$, on its negative side or lies 
on $H_{\alpha}$.

\par
Also notice that the line $x_1=x_2= ...=x_{N-1}=x_N$ is contained in every
plane $H_{\mbi\mbj}$, if the orthogonal complement to this line 
is $\mathcal{U} : x_1 + x_2 + ... + x_{N-1} + x_N = 0$, we can define 
a partition on $\mathcal{U}$, known to combinatorialists 
as $\mathcal{A}_{N \! - 1}$ [4-5], with the set of hyperplanes 
$\mathcal{H}_{\mbi\mbj} = \mathcal{U} \cap H_{\mbi\mbj}$.
For reasons that are explained below the points outside $\mathcal{U}$ 
are not relevant to our construction.

\par
The set of all points $x \ \epsilon \ \mathcal{U}$ having the same 
sign vector $S$ form a {\bf cell} in the decomposition of $\mathcal{U}$ 
induced by $\mathcal{A}_{N \! - 1}$, associated to this secomposition 
is the following important structure : the {\bf face poset}, which is the set 
of all cells induced by $\mathcal{A}_{N-1}$ ordered by inclusion. 
The maximal cells (all $(N \! \! - \! \! 1)$-dimensional) are called 
{\bf regions} and are shaped as polyhedral cones, the coordinates 
of the points in the interior of a region obey the relation :

\vskip 2mm
\noindent
\hspace*{10mm} $x_{i_1} < x_{i_2} < ... < x_{i_{N-1}} < x_{i_N}$
\hspace*{10mm} (1)

\vskip 2mm
\noindent
the dominance relations (1) between the coordinates can be encoded 
by the sequence

\vskip 2mm
\noindent
\hspace*{10mm} $(i_1) (i_2) \ ... \ (i_{N-1}) (i_N)$
\hspace*{23mm} (2)

thereafter referred as the cell {\bf dominance partition sequence} ($DPS$), 
where the set of indices $i_{\alpha}$ is a permutation 
of $(1,2,...,N \! - \! 1,N)$. Each index appears enclosed between parenthesis 
for reasons that will be made clear in the next section.

\par
Reflecting a point in general position on $\mathcal{H}_{\mbi\mbj}$ gives 
an image where the coordinates $\mbi$ and $\mbj$ are switched and  
the others are left unchanged. Multiple reflections of a point 
on the hyperplanes $\mathcal{H}_{\mbi\mbj}$ generate a set of $N!$ images 
which are the permutations of its coordinates. This leads to the fact 
that the $\binom{N}{2}$ hyperplanes form a Coxeter reflection 
arrangement [7] whose symmetry group is isomorphic to the
symmetric group $S_N$ of permutations of the set $(1,2,...,N \! - \! 1,N)$.

\par
The reflection group $\mathcal{A}_{N-1}$ is also the symmetry group 
of a polytope:
the {\bf $N$-permutohedron} or $\Pi_{N-1}$ [5], so called because 
its vertices are obtained by permuting the coordinates 
of the vector $(1,2,...,N \! - \! 1,N)$. The faces of the the permutohedron 
are polar to the cells of the hyperplane arrangement and the face lattices 
of both are isomorphic.

\par
For a molecule with $N$ atoms as the $x$, $y$ and $z$ coordinates 
are independent of each other [1] we have a $\mathcal{A}_{N-1}$ 
partition for each of them, that is ${\mathcal{A}_{N-1}}^{\text{\small{3}}}$ 
for the whole $CS$.
As it has been emphasized in [1-2] the $-1$ is because of the translation 
symmetry : the conformations outside the hyperplane $\mathcal{U}$ 
correspond to translated $3D$ structures.

\par
The radial dimension in $CS$ is also spurious: multiplying the coordinates 
of an arbitrary $3D$ conformation by a positive factor generates a set 
of points lying on a half-line starting at the origin. The partition 
$\mathcal{A}_{N-1}$ is {\it central} because that takes into account 
the scaling symmetry. 

\par
${\mathcal{A}_{N-1}}^{\text{\small{3}}}$ on the other hand does not take 
into account the rotation symmetry [2], the solution of this problem and
its consequences will be discussed in sections IV to VII. 

\begin{center}
{\scshape III. The face lattice poset}
\end{center}

\par
The combinatorial structure of the $\mathcal{A}_{N-1}$ face poset 
is the fundamental concept behind this work, it can be understood 
by studying a class of objects called {\bf tournaments} which are directed 
graphs with $\mathcal{N}$ nodes [8], these are used to investigate 
the properties of permutations, so useful for characterizing the cells in $CS$.

A permutation of a set of $\mathcal{N}$ elements can be represented 
by an acyclic, complete and labelled tournament (see fig. 1 for a description),
where :
\begin{itemize}
\item The term {\bf acyclic} means that the graph contains no directed cycles.
\item A graph is {\bf complete} if there is always an arc between 
      any two nodes, if an arc goes from $i$ to $j$ we say that $i$ 
      {\bf dominates} $j$. The {\bf score} of a node is the number of nodes 
      it dominates.
\item Each node of the graph has a unique {\bf label} which is a number 
      between 1 and $\mathcal{N}$ that distinguishes it from the other nodes.
\end{itemize}
In what follows the term tournament refers exclusively to tournaments 
where the above qualifiers apply.

\vskip 2mm
\includegraphics{JGAFigure.1}

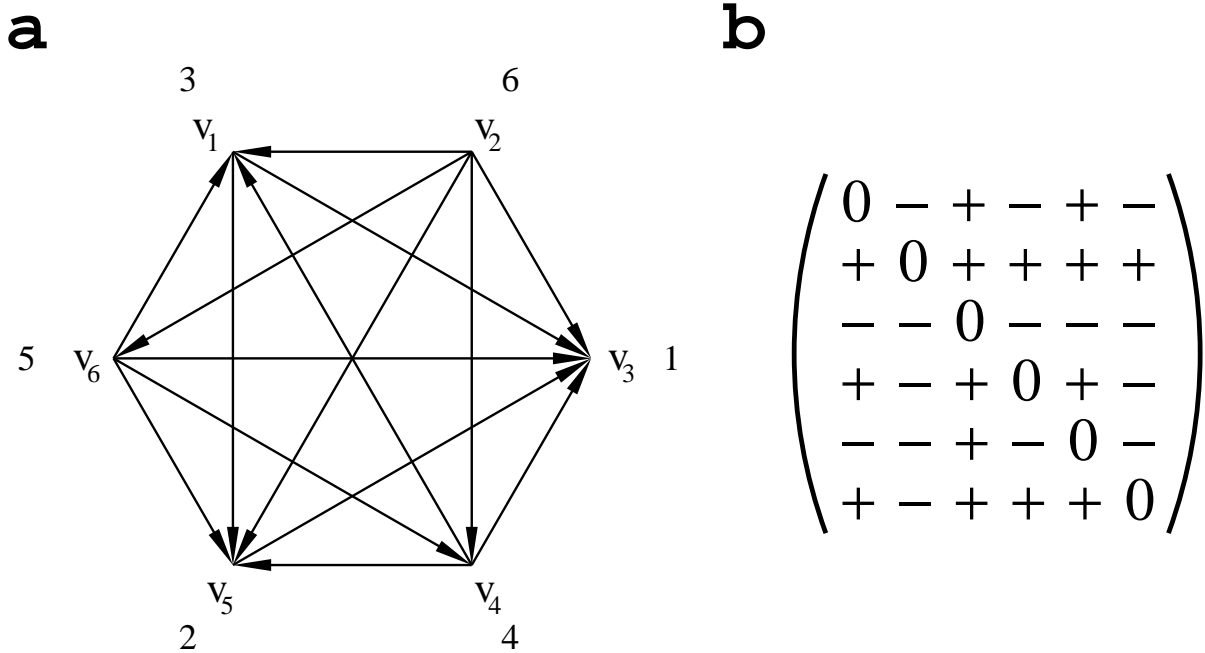
{\footnotesize Figure 1. 

a) A complete acyclic tournament corresponding to the permutation 
   $(3,6,1,4,2,5)$ which is the score of each vertex plus 1, the indices 
   in the dominance sequence of vertices 
   $ (3)(5)(1)(4)(6)(2) $ correspond to the inverse permutation. \newline
b) The antisymmetric incidence matrix, the rows in the upper triangle form 
   the sign vector.}

\vskip 2mm
For a tournament with $\mathcal{N}$ nodes the following statements are true :
\renewcommand{\theenumi}{\Roman{enumi}}
\begin{enumerate}
\item {\it In a tournament there is always a node called the {\bf sink} 
       that is dominated by every other node}.
       \newline
       Consider the last node of any maximal directed path, if an arc connects
       it to another node then either the path is not maximal or there is 
       a cycle; if there were another sink it would be connected to the
       former and either it would dominate or be dominated. 

\item {\it In a tournament there is always a node called the {\bf source} 
       that dominates every other node}.

\item {\it Any subgraph of a tournament is also a tournament}.
       \newline
       Any subgraph from a complete graph is also complete, and it can contain
       no cycles otherwise they would also be present in the parent graph.

\item {\it There is one maximal path that spans the graph}.
       \newline
       Consider the subtournament obtained by removing the source, then
       start the path with the arc that goes from the source to the subsource,
       and repeat the same step with the subgraph until you reach the sink.
       The path obtained goes through every node since 
       there are $\mathcal{N}-1$ steps, and is maximal since skipping a 
       subsource for another node shortens the path since the node is
       dominated by the subsource. 

\item {\it The sequence of labels of the nodes visited by the maximal path
        is the dominance partition sequence}.
       \newline
       By the construction procedure the first node, the source, dominates 
       all other nodes, the second dominates the remaining nodes and so on.

\end{enumerate}

{\bf Theorem 1}. {\it In a tournament the arcs between a set of consecutive 
                  nodes in the maximal path can be arbitrarily reversed and 
                  the resulting graph still be a tournament if the subgraph 
                  spanning the consecutive nodes is a tournament}.

\par
Since the subgraph and its complement are tournaments they contain
no cycles, thus a cycle must involve nodes between the subgraph and 
the complement, but this is not possible since by construction
the set of consecutive nodes is dominated by the preceeding nodes in the
maximal path and likewise it dominates the following ones.

\par
By V reversing an arc between contiguous nodes is equivalent 
to a transposition in the $DPS$. 

{\bf Theorem 2}. {\it In a tournament encoded by 
                  $(i_1) (i_2) \ ... \ (i_{\alpha}) 
                               \ ... \ (i_{\alpha+n-1}) 
                               \ ... \ (i_{N-1}) (i_N)$
                  the permutations in the set of $n$ consecutive indices
                  $i_{\alpha} \ ... \ i_{\alpha+n-1}$ give a set 
                  of tournaments that encode the vertices 
                  of an $n$-permutohedron}.

\par
If we restrict ourselves to the $n$-dimensional subspace spanned by the 
coordinates $(x_{i_{\alpha}}, ... ,x_{i_{\alpha+n-1}})$ the permutations
of the indices above corresponds to the permutations of the coordinates
of the vector $(\alpha,\alpha+1, ...,\alpha+n-1)$ which are the vertices
of a $\Pi_{n-1}$.

{\bf Corollary}. {\it The $n$-permutohedron is a face of $\Pi_{N-1}$}.

Obviously since it is contained in the affine hyperplane 
$x_{i_{\alpha}} + x_{i_{\alpha+1}} + ... + x_{i_{\alpha+n-1}} = 
 n (\alpha+(n-1)/2)$. This face is encoded by the $DPS$

\hspace*{10mm} 
$(i_1) (i_2) \ ... \ (i_{\alpha} 
             \ ... \  i_{\alpha+n-1}) \ ... \ (i_{N-1}) (i_N)$
\hspace*{10mm} (3)

\par
that represents the set of $n!$ sequences that are permutations 
of the indices $i_{\alpha}$ to $i_{\alpha+n-1}$.

{\bf Corollary}. {\it The sequence 
                  $(i_1) (i_2) \ ... \ (i_{\alpha} 
                               \ ... \  i_{\alpha+n-1})
                               \ ... \ (i_{\beta} 
                               \ ... \  i_{\beta+m-1}) 
                               \ ... \ (i_{N-1}) (i_N)$
                  encodes the $(n+m-2)$-face $\Pi_{n-1} \times \Pi_{m-1}$}.

\par
This can be seen from the definition given above of the product of polytopes.

\par
Thus the meaning of parenthesis in $DPS$s becomes apparent : 
each parenthesis enclosing a sequence of length $n$ encodes 
an $\Pi_{n-1}$ polytope, and the whole sequence encodes the product 
of all these polytopes. 

\par
These sequences can be ordered by inclusion to form a face lattice poset, 
which is isomorph to the one obtained with the sign vectors, since like 
$DPS$s they are another encoding scheme for tournaments [1]. 

\par
This is an important feature because it implies the {\bf modularity} 
of the model: the face lattice of a molecule can be obtained 
as the product of the face lattices of subsets of atoms.

\newpage
\begin{center}
{\scshape IV. Enumerating the orientations of a simplex}
\end{center}

For a simplex with random morphology we define the set of vectors that
run along the edges and their associated central planes (figs. 2a and 2b)

\vskip 2mm
\noindent
\hspace*{10mm} $ e_{\mbi\mbj}=v_{\mbi}-v_{\mbj} $ \hspace*{1mm}, 
               $ 1 \leq \mbi < \mbj \leq 4  $     \hspace*{10mm} (4)

\hspace*{10mm} $ \mathcal{E}_{\mbi\mbj}^0(x)=\{x \ \epsilon \ \realset^3 : 
                 e_{\mbi\mbj}.x=0\} $ \hspace*{10mm} (5)

\vskip 2mm
\noindent
Each plane divides $3D$ space into positive and negative halves

\vskip 2mm
\noindent
\hspace*{10mm} $\mathcal{E}_{\mbi\mbj}^+(x)=\{x \ \epsilon \ \realset^3 :
                e_{\mbi\mbj}.x > 0\}$ \ \ and  \ \
               $\mathcal{E}_{\mbi\mbj}^-(x)=\{x \ \epsilon \ \realset^3 :
                e_{\mbi\mbj}.x < 0\}$ \ \ \ \ \ \ \ \ \ (6)

\par
As for the regular tetrahedron described above (5) and (6) generate an
$\mathcal{A}_3$ partition of $3D$ space in 24 irregular shaped 
cells, fig. 2b.

\vskip 2mm
\includegraphics{JGAFigure.2}

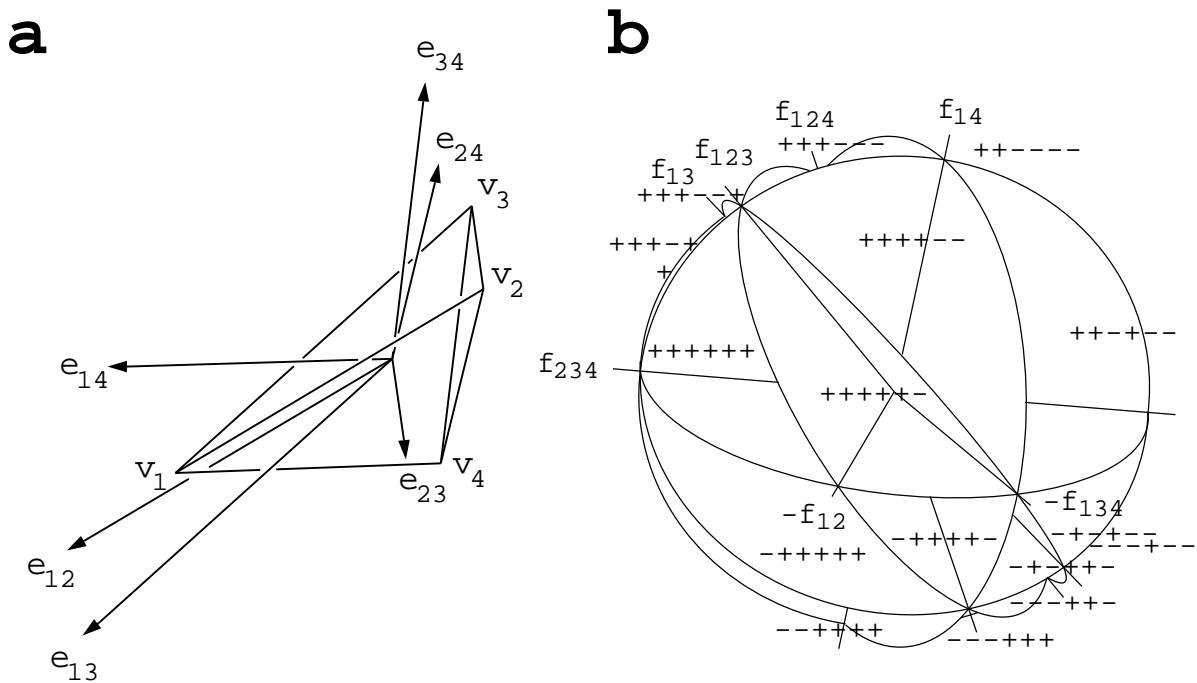
{\footnotesize Figure 2. The $\mathcal{A}_3$ partition of a random simplex.

a) The random simplex with the vectors $e_{\mbi\mbj}$ 
   centered at the origin. \newline
b) The partition of $3D$-space by the planes $\mathcal{E}_{\mbi\mbj}$ 
   represented as intersecting disks centered at the origin, visible 
   $3D$ cells are designated by their sign vector and $1$-dimensional 
   cells are labelled by the corresponding $f_{\text{...}}$ symbols (7).}

\par
This partition has the following interesting property: assume for instance 
that the $x$ axis of a central orthogonal reference system in general position 
lies entirely within the cell encoded by the permutation $(3,1,4,2)$, 
or equivalently the sign vector $(+-+--+)$, then the dominance relation 
$v_{2_x} < v_{4_x} < v_{1_x} < v_{3_x} $ holds for the $x$ coordinates 
of the vertices of the simplex.

\par
This suggests a method for enumerating the cells 
in ${\mathcal{A}_3}^{\text{\small{3}}}$ that correspond to the different 
orientations\footnote{All along this work the term {\bf orientation} is
used interchangeably with $DPS$ and sign vector.}
of the simplex : it suffices to enumerate the cells with 
the lowest dimensions, the more numerous $(3,3,3)$-dimensional cells 
can be easily obtained through the connecting paths in the face lattice.

\par
The $1$-dimensional cells in ${\mathcal{A}_3}$ are determined by the set 
of vectors perpendicular to the faces of the simplex and to pairs of opposite 
edges

\vskip 2mm
\noindent
\hspace*{10mm} $ f_{123}=e_{12} \wedge e_{23} \ , \
                 f_{124}=e_{12} \wedge e_{24} \ , \
                 f_{134}=e_{13} \wedge e_{34} \ , \
                 f_{234}=e_{23} \wedge e_{34} \ , $  \newline
\hspace*{10mm} $ f_{12}=e_{12} \wedge e_{34} \ , \
                 f_{13}=e_{13} \wedge e_{24} \ , \
                 f_{14}=e_{14} \wedge e_{23} $ 
\hspace*{40mm}  (7)

\vskip 2mm
their corresponding central planes will be designated 
$\mathcal{F}_{\mbi\mbj\mbk}$ and $\mathcal{F}_{\mbi\mbj}$.

\par
If we take the sign of the scalar products between the sets of vectors (4) 
and (7) we obtain a matrix

\vskip 2mm
\noindent
\hspace*{10mm}
$\begin{matrix}
          & e_{12} & e_{13} & e_{14} & e_{23} & e_{24} & e_{34} & DPS      \\
f_{123}   &  0     &  0     &  +     &  0     &  +     &  +     & (4)(123) \\
f_{124}   &  0     &  -     &  0     &  -     &  0     &  +     & (234)(3) \\
f_{134}   &  +     &  0     &  0     &  -     &  -     &  0     & (2)(134) \\
f_{234}   &  +     &  +     &  +     &  0     &  0     &  0     & (234)(1) \\
f_{12}    &  0     &  -     &  -     &  -     &  -     &  0     & (12)(34) \\
f_{13}    &  +     &  0     &  +     &  -     &  0     &  +     & (24)(13) \\
f_{14}    &  -     &  -     &  0     &  0     &  +     &  +     & (14)(23)
\end{matrix}$ \hspace*{47.5mm}  (8)

\vskip 2mm
\par
that up to a sign reversal is an invariant [4,9], it is the same for 
any simplex whatever its morphology. The rows are the sign vectors 
of the $1$-dimensional cells with the corresponding dominance partition 
sequence on the righ, these cells can be seen in fig. 2b where the labels
$f_{\mbi\mbj\mbk}$ and $f_{\mbi\mbj}$ are on top of the lines intersected 
by the planes $\mathcal{E}_{\mbi\mbj}$, $\mathcal{E}_{\mbi\mbk}$, 
$\mathcal{E}_{\mbj\mbk}$ and $\mathcal{E}_{\mbi\mbj}$, 
$\mathcal{E}_{\mbk\mbl}$ respectively.

\par
We start by enumerating the orientations of a reference system whose
$z$ axis is parallel to one of the vectors (7), $f_{123}$ for example,
the remaining axis $x$ and $y$ will be on the plane $\mathcal{F}_{123}$,
the problem is to determine how  the $\mathcal{E}_{\mbi\mbj}$s (5) divide 
this plane into $2$-dimensional cells. In fig. 3 we can see 
the four possible 12-sector partitions that can be
generated by the vectors $e_{12}$, $e_{13}$ and $e_{23}$ and the
perpendicular intersections of the planes $\mathcal{E}_{12}$, 
$\mathcal{E}_{13}$ and $\mathcal{E}_{23}$. This partition gives us
only half of the sign vectors components, to obtain the remaining ones
we need to introduce a morphological classification of simplexes.

\begin{center}
{\scshape V. Morphological classification of simplexes }
\end{center}

\par
For a given simplex, like the one in fig. 2a for instance, we compute the sign 
of the scalar products of the vectors (4) and (7) between them, this gives 
the following two tables

\vskip 2mm
\noindent\hspace*{10mm}
$\begin{matrix}
       & e_{13} & e_{14} & e_{23} & e_{24} & e_{34} \\
e_{12} &    +   &    +   &    -   &    -   &    -   \\
e_{13} &        &    +   &    +   &    -   &    -   \\
e_{14} &        &        &    +   &    +   &    +   \\
e_{23} &        &        &        &    +   &    +   \\
e_{24} &        &        &        &        &    +
\end{matrix}$ \hspace*{5mm} 
$\begin{matrix}
        & f_{124} & f_{134} & f_{234} & f_{12}  & f_{13} & f_{14} \\
f_{123} &    +    &    +    &    +    &    -    &    +   &    +   \\
f_{124} &         &    +    &    +    &    +    &    +   &    +   \\
f_{134} &         &         &    +    &    +    &    +   &    -   \\
f_{234} &         &         &         &    -    &    +   &    -   \\
f_{12}  &         &         &         &         &    +   &    +   \\
f_{13}  &         &         &         &         &        &    +
\end{matrix}$ \hspace*{4.5mm}  (9)

\vskip 2mm
\par
The set of signs (9) refer mostly to angles between adjacent edges and
dihedral angles between contiguous faces: $+$, $0$ and $-$ are for acute, 
right and obtuse angles respectively. 

\newpage
\includegraphics{JGAFigure.3}

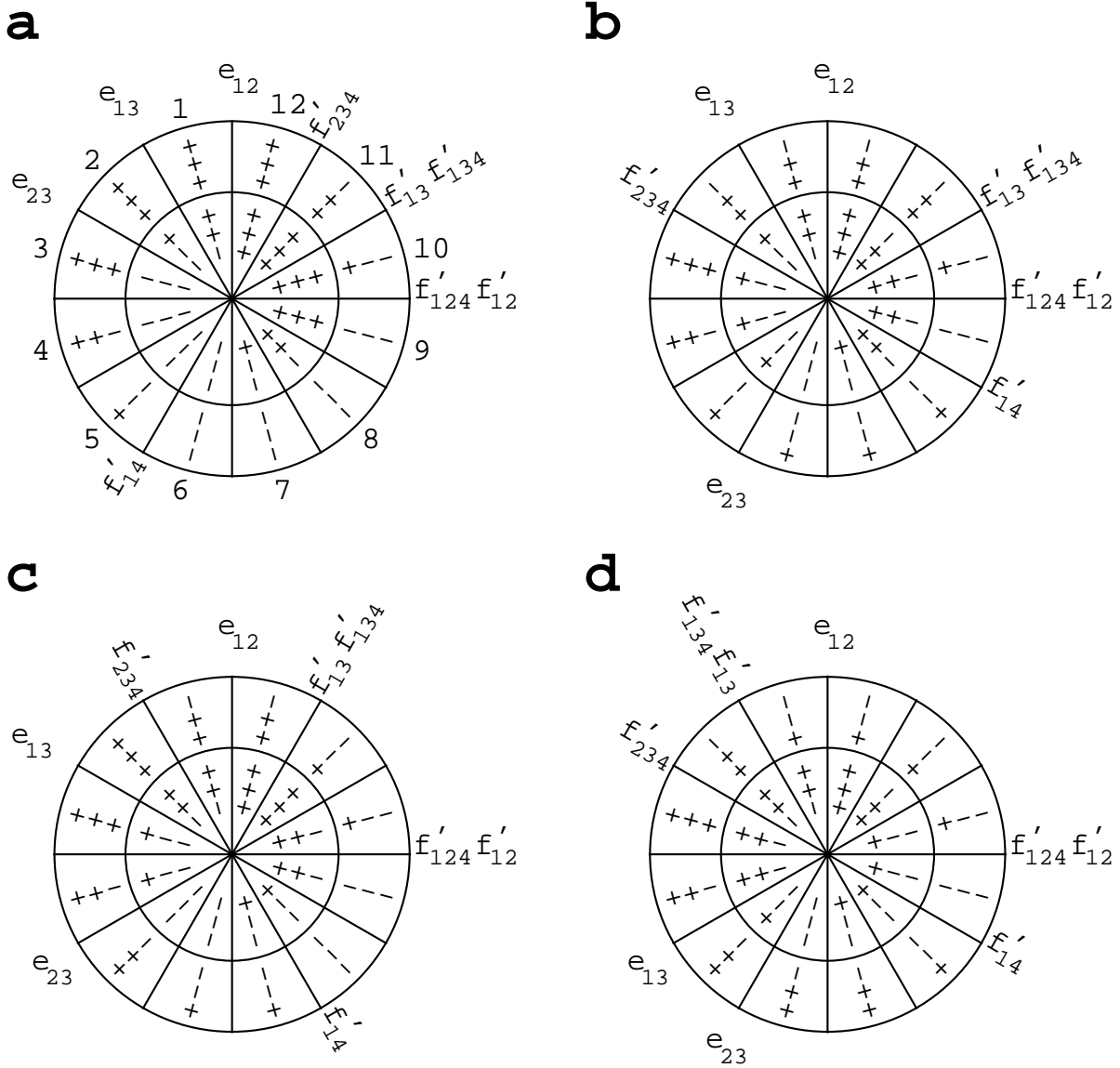
{\footnotesize Figure 3. The four possible partitions of the plane 
$\mathcal{F}_{123}$. 

Within figs. {\bf a} to  {\bf d} the vector $f_{123}$ points 
in the upward direction, the labels $e_{12}$, $e_{13}$ and $e_{23}$ 
are over the lines that run along these vectors, and the corresponding 
perpendicular lines are the intersections with the planes $\mathcal{E}_{12}$, 
$\mathcal{E}_{13}$ and $\mathcal{E}_{23}$ respectively.
The label $f_{\mbi\mbj\mbk}^{'}$ means that the corresponding line runs along
the projection of vector $f_{\mbi\mbj\mbk}$ on $\mathcal{F}_{123}$. 
The labels $f_{124}^{'}$ and $f_{12}^{'}$ over the intersection 
of plane $\mathcal{E}_{12}$, for instance, is because  $f_{124}$ and $f_{12}$
are contained in that plane, and reciprocally $e_{12}$ is contained 
in the planes $\mathcal{F}_{124}$ and $\mathcal{F}_{12}$. \newline
All these lines converge at the origin and partition $\mathcal{F}_{123}$ 
in 12 sectors : between the inner and outer circles are the sign vector
components of $e_{12}$, $e_{13}$ and $e_{23}$, for each sector they should
be read from inside out in that order; within the inner circle 
there are the sign vector components of $f_{124}^{'}$, $f_{134}^{'}$ 
and $f_{234}^{'}$ respectively. \newline
The sectors are numbered from 1 to 12 as indicated in {\bf a}.}

\par
Thus the rough morphological characteristics of a simplex can be encoded 
in a 36 bit binary\footnote{We exclude sequences harboring $0$s as they form 
a set of null measure. }
sequence : there are a total of 3936 sequences that correspond 
to geometrically realizable simplexes, these define the set of morphological 
classes {\it {\bf A}} of labelled simplexes. We define the {\bf volume} 
of a class as the set of cells it spans 
in ${\mathcal{A}_3}^{\text{\small{3}}}$.

\par
It should be reminded that this classification has a graph structure, 
since geometrical deformations in a simplex from one class induce a transition 
to other classes thus establishing a connectivity between them; the precise 
structure of such a graph is of no utility in the present work, 
but the concept is important when we will introduce below the dynamical 
states of a simplex.

\par
The binary sequence (9) is instrumental in finding the partition of the planes
perpendicular to $1$-dimensional cells, in our exemple it can be deduced from
(9) that the partition of $\mathcal{F}_{123}$ is the one of fig. 3c, since it
is the only one that satisfies the relation

\noindent\hspace*{10mm}
$(SIGN(e_{12}.e_{13}),SIGN(e_{12}.e_{23}),SIGN(e_{13}.e_{23})) = (+ - +)$

There are also the relations concerning vectors  $e_{14}$, $e_{24}$ 
and $e_{34}$

\noindent\hspace*{10mm}
$(SIGN(e_{14}.e_{12}),SIGN(e_{14}.e_{13}),SIGN(e_{14}.e_{23})) = 
(+ + +)$ \hspace*{14mm} (10a) \\
\noindent\hspace*{10mm}
$(SIGN(e_{24}.e_{12}),SIGN(e_{24}.e_{13}),SIGN(e_{24}.e_{23})) = 
(- - +)$ \hspace*{14mm} (10b) \\
\noindent\hspace*{10mm}
$(SIGN(e_{34}.e_{12}),SIGN(e_{34}.e_{13}),SIGN(e_{34}.e_{23})) = 
(- - +)$ \hspace*{14mm} (10c)

thus $e_{14}^{'}$, the projection\footnote{The $'$ superscript designates 
the projection of a vector on $\mathcal{F}_{...}$.}
of $e_{14}$, must lie in sectors 2 or 3 by (10a); similarly $e_{24}^{'}$ 
and $e_{34}^{'}$ must be in sectors 6 or 7 by (10b) and (10c). 
These ambiguities can be resolved by set of relations 

\noindent\hspace*{10mm}
$(SIGN(e_{14}.f_{123}),SIGN(e_{24}.f_{123}),SIGN(e_{34}.f_{123}))$ 
\hspace*{3.4mm} $= (+ + +)$ \hspace*{5mm} (11a) \\
\noindent\hspace*{10mm} 
$(SIGN(f_{124}.f_{123}),SIGN(f_{134}.f_{123}),SIGN(f_{234}.f_{123}))$ 
                $= (+ + +)$ \hspace*{5mm} (11b) \\
\noindent\hspace*{10mm} 
$(SIGN(f_{12}.f_{123}),SIGN(f_{13}.f_{123}),SIGN(f_{14}.f_{123}))$ 
\hspace*{3.4mm} $= (- + +)$ \hspace*{5mm} (11c)

$e_{14}$ for instance, lies on $\mathcal{F}_{124}$ and together with $f_{124}$ 
stands above $\mathcal{F}_{123}$, by (11a) and (11b), this implies that 
$SIGN(e_{14}^{'}.f_{124}^{'}) = -$. Repeating this procedure for $f_{134}$ and
$f_{234}$, and for each of the vectors $e_{24}$ and $e_{34}$ we end up with

\noindent\hspace*{10mm}
$(SIGN(e_{14}^{'}.f_{124}^{'}),SIGN(e_{14}^{'}.f_{134}^{'}),
  SIGN(e_{14}^{'}.f_{14}^{'}))  = (- - -)$  \hspace*{11.5mm} (12a) \\
\noindent\hspace*{10mm}
$(SIGN(e_{24}^{'}.f_{124}^{'}),SIGN(e_{24}^{'}.f_{13}^{'}),
  SIGN(e_{24}^{'}.f_{234}^{'})) = (- - -)$  \hspace*{11.5mm} (12b) \\
\noindent\hspace*{10mm}
$(SIGN(e_{34}^{'}.f_{12}^{'}),SIGN(e_{34}^{'}.f_{134}^{'}),
  SIGN(e_{34}^{'}.f_{234}^{'})) = (+ - -)$  \hspace*{11.5mm} (12c)

(12a), (12b) and (12c) imply that $e_{14}^{'}$, $e_{24}^{'}$ and $e_{34}^{'}$
are to be found in sectors 3, 6 and 7 respectively, thus removing 
these ambiguities.

\par
There is one ambiguity though that cannot be resolved by the binary 
sequence (9) : 
$\mathcal{H}_{24}$ runs through sectors 3 and 9 together with $e_{14}^{'}$,
and $\mathcal{H}_{14}$ runs through sectors 6 and 12 as $e_{24}^{'}$ , 
so we end up with two possible partitions of $\mathcal{F}_{123}$ 
that are shown in fig. 4.

\par
As can be seen from fig. 4 each partition generates 12 $2$-dimensional cells 
and the same number in one dimension, by construction the lines along 
the $1$-dimensional cells are never perpendicular to each other, as a
consequence for an $(x,y)$ reference system centered at the origin if one 
of the axis runs along the edge of a sector the other will be located inside 
a sector: rotating the axis system enables us to scan 
12 $(1,2,3)$ and 12 $(2,1,3)$ dimensional cells (see fig. 4).

\par
Thus for any orientation structure associated with a plane 
$\mathcal{F}_{ ... }$, a reference system with one axis 
perpendicular to the plane can be in $2 \times 12 \times 6$ cells 
with dimensions any permutation of the sequence $(3,2,1)$ in $(x,y,z)$.
This solves the problem of enumerating the cells with the lowest possible
dimensions that correspond to an orientation of the simplex, 
the $(3,3,3)$-dimensional cells can be found from these through the connecting 
paths in the ${\mathcal{A}_3}^{\text{\small{3}}}$ cell lattice poset.

\includegraphics{JGAFigure.4}

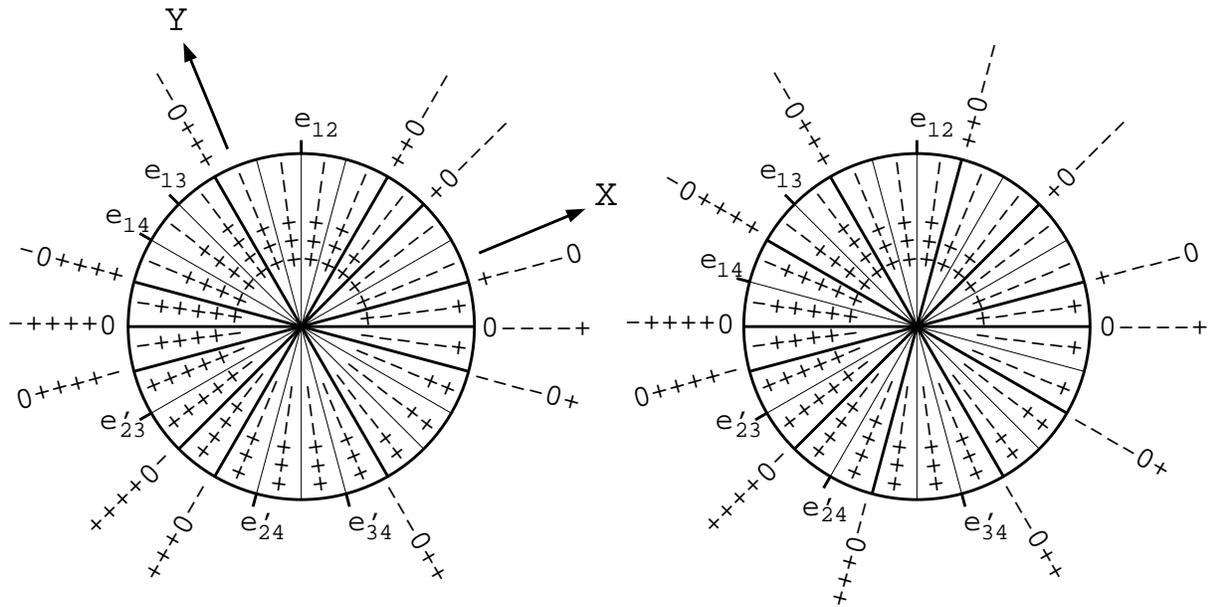
{\footnotesize Figure 4. The two possible orientation structures
of $\mathcal{F}_{123}$.

The thick lines are the intersections of $\mathcal{E}_{\mbi\mbj}$ 
with $\mathcal{F}_{123}$, the thin ones are lines along the vectors 
$e_{\mbi\mbj}$ and $e_{\mbi\mbj}^{'}$.
The sign vectors of the $2$-dimensional cells lie inside the circle, 
the $1$-dimensional ones are outside along the corresponding partition line,
they should be read from inside out. 
An (X,Y) axis system has been superimposed on the first structure 
as a visual aid to show how the sectors can be scanned.}

\par
Thus for any orientation structure associated with a plane 
$\mathcal{F}_{ ... }$, a reference system with one axis 
perpendicular to the plane can be in $2 \times 12 \times 6$ cells 
with dimensions any permutation of the sequence $(3,2,1)$ in $(x,y,z)$.
This solves the problem of enumerating the cells with the lowest possible
dimensions that correspond to an orientation of the simplex, 
the $(3,3,3)$-dimensional cells can be found from these through the connecting 
paths in the ${\mathcal{A}_3}^{\text{\small{3}}}$ cell lattice poset.

\begin{center}
{\scshape VI. The conformational space of a simplex}
\end{center}

We have seen that the binary sequences (9) cannot define unambiguous 
partitions of the planes $\mathcal{F}_{...}$ : for 
each $\mathcal{F}_{\mbi\mbj\mbk}$ there can be between 1 and 3 possible 
orientation structures, and between 1 and 24 
for each $\mathcal{F}_{\mbi\mbj}$; in a given class only a fraction 
of the combinations between the different orientation structures, 
one from each plane, give geometrically realizable simplexes.

\par
To remove ambiguities we need to define a set {\it {\bf B}} of morphological 
classes such that for each one the range of geometrical 
variation only allows one orientation structure per $\mathcal{F}_{...}$. 
An empirical Monte Carlo calculation yields a total of 125712 classes 
of labelled simplexes, a class {\it {\bf A}} has a number of subclasses 
{\it {\bf B}} that goes from a minimum of 1 up to a maximum of 220. 
These morphological subclasses have the remarkable property 
that for {\it a $3D$ conformation any cell in the volume can be reached 
through a rotation}, which is is an obvious consequence of the one 
to one correspondence between $\mathcal{F}_{...}$ planes and orientation 
structures.

\par
Thus a class {\it {\bf A}} can be decomposed into a set of subclasses 
{\it {\bf B}}, that can be unambiguously oriented in a standard $3D$
reference frame, and its volume in $CS$ is simply the union of the volumes 
of its subclasses.

\newpage
\begin{center}
{\scshape VII. The orientation structures}
\end{center}

\par
To achieve a morphological classification of simplexes we need to know
how many classes of orientation structures there are, since the classes 
{\it {\bf A}} decompose into subclasses {\it {\bf B}} and each of these 
is determined by 7 orientation structures.

\par
A first classification concerns the circular order of the vectors 
$e_{\mbi\mbj}^{'}$ in the plane $\mathcal{F}_{\alpha}$. This can be deduced 
from the set of signs (9), for instance : 
by (4) and (7) the shortest circular path going through $e_{12}^{'}$, 
$e_{13}^{'}$ and $e_{23}^{'}$ must be less than $\pi$, and it runs
counter-clockwise if the sign of $\mathcal{F}_{123}.\mathcal{F}_{\alpha}$ 
is $+$.

\par
This exemple leads to the general solution that was discussed in [2] : 
the 7 vectors (7) define a central partition dual to $\mathcal{A}_3$ [9] 
that divides the $3D$ space in 32 cells.
The sign vector of the cell that contains $\mathcal{F}_{\alpha}$
defines the sense of the shortest circular path that connects 
the projected vectors in the 7 ordered sets 
$\{e_{12}^{'}, e_{13}^{'}, e_{23}^{'}\}$,
$\{e_{12}^{'}, e_{14}^{'}, e_{24}^{'}\}$,
$\{e_{13}^{'}, e_{14}^{'}, e_{34}^{'}\}$,
$\{e_{23}^{'}, e_{24}^{'}, e_{34}^{'}\}$,
$\{e_{12}^{'}, e_{34}^{'}\}$,
$\{e_{13}^{'}, e_{24}^{'}\}$ and
$\{e_{14}^{'}, e_{23}^{'}\}$. This generates a set of 7 constraints from
which the circular order of the $e_{\mbi\mbj}^{'}$s in $\mathcal{F}_{\alpha}$ 
can be deduced, making a total of 32 possible circular orientations.

\par
As can be seen in fig. 4 on the plane $\mathcal{F}_{\alpha}$ 
each $e_{\mbi\mbj}^{'}$ contributes a total of 4 separations between sectors 
at periodic intervals of $90^{\text{o}}$ each comprising exactly 6 sectors, 
on the other hand there are two classes of separations : 
either a line along the vector $e_{\mbi\mbj}^{'}$
or the intersection of a plane $\mathcal{H}_{\mbi\mbj}$, in an interval 
of $90^{\text{o}}$ the possible distributions of the two separators amounts 
to a total of $2^{\text{\small{5}}}$ combinations.
This makes 1024 classes of orientation structures like those in fig. 4, 
among these 48 appear to be not geometrically realizable 
since they are not found in any class {\it {\bf B}}.

\begin{center}
{\scshape VIII. Determination of the graph of cells}
\end{center}

\par
Most often in mesoscopic models of biological macromolecules atoms are
represented as point-like structures surrounded by an atomic force 
field [10,11], thus any four atoms are the vertices of a 3-simplex.
Also for a molecular system with $N$ atoms an order relation can be defined
by numbering its atoms from $1$ to $N$, so that 3-simplexes can be designated
as a $4$-tuple of ordered integers which are the numbers of its atoms.

\par
Beyond the orientation problem, the classes {\it {\bf A}} and {\it {\bf B}} 
bring the possibility of analizing the dynamics of a molecular system in terms 
of discrete entities, the range of morphological variation for simplexes 
within a molecule can be explored in molecular dynamics simulations ($MDS$) 
and the results can be summarized as follows [2,12]
\begin{itemize}
\item 90\% of simplexes in a structure evolve within less 
      than 20 classes {\it {\bf A}}.
\item The maximum variation observed is somewhat less than 200 classes, about
      5\% of the total.
\end{itemize}

\par
This result opens up the possibility of determining the set of geometrically 
accesible cells in the $CS$ of a molecular system.

\par
The $CS$ of a simplex has a total of 13824 cells and, typically, the volume 
of a class {\it {\bf A}} is about one third of that number, much less 
if we exclude structures that can be derived through a rotation. 
This volume is very small when compared to the huge number of cells spanned 
by a molecular system, and it can be  reasonably assumed that the volume
of a simplex can be scanned by a molecular dynamics run. 
What cannot be scanned by a simulation is the set of structures that arise
by combining the local movements.

\par
MDSs can be used to determine the subgraph of classes spanned
by every simplex, and the volume of the molecular system in $CS$ can be
obtained by progressively merging the $CS$ of individual simplexes. 
As we were able to determine the different orientations 
of a simplex this process can be done excluding redundant rotated structures.

\par
Before proceeding further let us show with a simple exemple the basic 
operations that are involved in the process of merging $CS$s.
If we have two adjoining simplexes $S_{\alpha}$ and $S_{\beta}$ 
represented by the tetrads $\{14,33,82,86\}$ and $\{14,82,86,91\}$ 
respectively (notice that their common faces correspond to the vertices 
$(v_1,v_3,v_4)$ and $(v_1,v_2,v_3)$), 
if the $3D$ structure of $S_{\alpha}$ is in a cell encoded by 
the  dominance partition sequence

\noindent\hspace*{10mm}
$((82)(14)(86)(33), (33)(82)(86)(14), (86)(14)(33)(82))$ \hspace*{26.5mm} (13)

then the set cells in $CS_{\beta}$ geometrically compatible with (13)
will be those whose $DPS$ contains the pattern 

\noindent\hspace*{10mm}
$((82)(14)(86), (82)(86)(14), (86)(14)(82))$ \hspace*{47.2mm} (14)

Thus a cell in $CS_{\beta}$ with $DPS$

\noindent\hspace*{10mm}
$((82)(91)(14)(86), (91)(82)(86)(14), (86)(14)(91)(82))$ \hspace*{27mm} (15)

can be merged with (13) and generates the set of $4$ cells 
in $CS_{\alpha} \times CS_{\beta}$

\noindent\hspace*{10mm}
$((82)(91)(14)(86)(33), (33 \ 91)(82)(86)(14), (86)(14)(33 \ 91)(82))$ 
\hspace*{10mm} (16)

which corresponds to a square face in the polar polytope.

\par
To calculate the graph of the geometrically accesible cells 
we begin by picking an arbitrary reference simplex, preferably one with low 
morphological variation, and arbitrarily choose an orientation among those 
available, this will be the simplex on level 1, the simplexes adjacent to this 
one form the level 2, and so on. Since adjacent simplexes in a $3D$ structure 
share three vertices the shortest adjacency path between any two of them  
has at most length 4, so we end up with simplexes in 5 levels. 

\par
We need not to include every simplex from the molecule to perform a useful
calculation, but there is the minimum requirement that every pair of atoms 
from a total of $\binom{N}{2}$ should be present at least once 
in a $4$-tuple, otherwise the $DPS$s could not be determined.

\par
The calculation can be done through the following procedure :
\renewcommand{\theenumi}{\arabic{enumi}}
\begin{enumerate}
\item Start at level 1.
\item From any simplex in level $n$ we select the compatible 
      orientations in the adjoining simplexes in level $n+1$.
\item From any simplex in the level $n+1$ we select compatible orientations 
      on the adjoining simplexes at the same level.
\item If $n < 5$ we go to step 2 and continue with level $n+1$.
\end{enumerate}
A link is created between any two compatible orientations in adjacent
simplexes. This is done in two steps:
\begin{enumerate}
\item If the simplex in the lower level has not yet been visited any 
      orientation compatible with those from the simplex in the upper level 
      is selected.
\item Otherwise any orientation that has not been selected is discarded.
      And likewise an orientation that fails to form a link with an adjacent 
      simplex is discarded because of geometrical inconsistency. 
\end{enumerate}

\par
The implementation of this procedure as an efficient computer algoritm
requires that the $CS$ of a class {\it {\bf A}} simplex be quickly 
searched for orientations compatible with those from the adjoining simplexes, 
these can be obtained from the set of orientation structures available 
to each $1$-dimensional cell $\mathcal{F}_{...}$ (7). 
This requirement can be fulfilled by building a hash table from where 
the $DPS$s like (15) can be retrived, such table has the following set 
of entries :
\begin{enumerate}
\item the number of the orientation class : from 1 to 976,
\item the connecting face, numbered from 1 to 4,
\item the $1$-dimensional $\mathcal{F}_{...}$ cell (7) corresponding 
      to the orientation structure, numbered from 1 to 7,
\item the chirality of the simplex: right or left-handed,
\item the pattern (14), of a total of 216 possible patterns.
\end{enumerate}

\begin{center}
{\scshape IX. Conclusion}
\end{center}

\par
The aim of the present work has been to bring the sheer complexity 
of molecular conformational space to tractable dimensions, by building 
a structure that encodes the set of geometrically accesible $3D$-conformations 
of a thermalized molecule, and putting it in a compact and manageable code.
The price to pay to achieve this result is the loss of the absolute precision 
over the local $3D$-conformations of molecular structures [1], but this has
no concern with this work since we only seek to obtain a global view of
conformational space. From this point of view the present formalism may be
a useful complement of molecular dynamics simulations that in the detailed
exploration of small regions is unexcelled.
\par
What remains to be done is to explore the graph of cells with a 
Hamiltonian functional over a force field and perform energy optimizations.
It should be emphasized that as a Hamiltonian is a function of distances
between atoms the present structure offers the possibility of calculating
the energy over entire regions of $CS$, since the interatomic distances can be
enumerated for a set of cells and in this case the energy function is 
nothing else than an integral over a rational function.

\begin{center}
{\scshape References}
\end{center}
\begin{itemize}

\item[[1]] J. Gabarro-Arpa, "A central partition of molecular conformational
           space. I. Basic structures" {\it Comp. Biol. and Chem., } 
           {\bf 27}, 153-159, (2003).

\item[[2]] J. Gabarro-Arpa, "A central partition of molecular conformational
           space. II. Embedding 3D-structures", {\it Proceedings 
           of the 26th Annual International Conference of the IEEE EMBS},
           San Francisco, 3007-3010 (2004).

\item[[3]] M. Karplus and J.A. McCammon, "Molecular dynamics simulations
           of biomolecules", 
           {\it Nature Struct. Biol. } {\bf 9}, 949-852 (2002).

\item[[4]] A. Bjorner, M. las Vergnas, B. Sturmfels, N. White, "Oriented 
           Matroids". Cambridge, UK, Cambridge University Press, 
           sect. {\bf 2} (1993).

\item[[5]] S. Fomin and N. Reading, 
           "Root systems and generalized associahedra",
           math.CO/0505518 (2005).

\item[[6]] G. Kreweras, "Sur les partitions non crois\'ees d'un cycle", 
           {\it Disc. Math.} {\bf 1}, 333-350 (1972).

\item[[7]] H.S.M. Coxeter, "Regular polytopes". Dover Publicaions, Inc., 
           New York (1973).

\item[[8]] J. W. Moon, "Topics on Tournaments". Holt, Rinehart and Winston, 
           New York (1968).

\item[[9]] J. Folkman, J. Lawrence, "Oriented matroids", {\it J. Combinatorial
           Theory B} {\bf 25}, 199-236 (1978).

\item[[10]] A.D. MacKerell Jr., et al., "All-Atom empirical potential 
            for molecular modeling and dynamics studies of proteins",
            {\it J. Phys. Chem. B} {\bf 102}, 3586-3616 (1998).

\item[[11]]  W. Wang, O. Donini, C.M. Reyes, P.A. Kollman,
             "Biomolecular simulations: recent developments in force fields, 
              simulations of enzyme catalysis, protein-ligand, 
              protein-protein, and protein-nucleic acid noncovalent 
              interactions",
             {\it Annu. Rev. Biophys. Biomol. Struct.} 
             {\bf 30}, 211-243 (2001).

\item[[12]] C. Laboulais, M. Ouali, M. Le Bret, J. Gabarro-Arpa, "Hamming
            distance geometry of a protein conformational space",
            {\it Proteins: Struct. Funct. Genet. } {\bf 47}, 169-179 (2002).

\end{itemize}

\end{document}